# Fines Migration and Permeability Decline during Reservoir Depletion Coupled with Clay Swelling due to Low-Salinity Water Injection: An Analytical Study

Suparit Tangparitkul[1,*], Alexander Saul[2], Cheowchan Leelasukseree[1],

Muhammad Yusuf[3] and Azim Kalantariasl[4]

[1]Department of Mining and Petroleum Engineering, Faculty of Engineering, Chiang Mai University, Chiang Mai 50200, Thailand

[2]School of Mechanical Engineering, University of Leeds, Leeds LS2 9JT, United Kingdom

[3]School of Chemical and Process Engineering, University of Leeds, Leeds LS2 9JT, United Kingdom

[4]Department of Petroleum Engineering, School of Chemical and Petroleum Engineering, Shiraz University, Shiraz, 71946-84471, Iran

[*]Corresponding author: suparit.t@cmu.ac.th  Tel: +66 5394 4128; Fax: +66 5394 4186

## Highlights

•   Particle retention has been modified to include effective stress and salinity dependents.

•   Reservoir configurations deform and particle detachment escalates due to reservoir depletion.

•   Clay swelling promotes blocking of flow paths leading to permeability decline.

# Abstract

Fines migration is crucial for reservoir permeability, involving both fine particle detachment and re-deposition along the flow channels. Fines migration behavior can either promote or obstruct fluid flow within the reservoir and is crucial for productivity optimization that needs fundamental understanding. The present work focuses on the contributions from effective stress build-up due to reservoir depletion and decreases in brine salinity from low-salinity water injection. Particle detachment is studied analytically by using the critical retention concentration function modified with stress and salinity dependents. Increase in formation effective stress leads to deformations within the reservoir configurations due to micro-cracks and reduced pore dimensions. Decreased size of the travelling channels promotes particle detachment, while the critical retention concentration decreases. A sensitivity study reveals that, under influence of effective stress, particle size and fluid velocity are dominant parameters controlling the fines migration by influencing the particle detaching forces. With decreasing brine salinity (*e.g.* via low-salinity water injection), clay particles that are attached on the pore surface increasingly swell, leading to reduction in effective pore space and flow channel. Decreased pore space directly obstructs travelling of suspended particles and fluids which results in permeability decline. Initial clay particle size is found to be a critical factor controlling particle detachment behavior. Small initial particle sizes are not likely to detach from the pore surface as a result of weak detaching forces, hence the analysis finds the critical retention concentration unchanged resulting in negligible permeability decline as a function of brine salinity. Considering both effects, effective stress and brine salinity, stress-dependent effects dominate salinity-dependent effects at low effective stress while the two contribute equally at high effective stress. Permeability decline is also determined from analytical results of the fine particle critical retention concentration, which emphasizes the roles of effective stress and brine salinity on the flow in reservoir porous media.

***Keywords:*** fines migration; permeability decline; effective stress; reservoir depletion; low-salinity waterflooding; fines-assisted waterflood

# 1 Introduction

Fine particle transportation in porous media (*e.g.* petroleum reservoir and ground water reservoir) has been studied over decades due to its importance in controlling and optimizing the fluid flow in pore or capillary channels.[1-7] It has been well-documented that the related mechanisms include particle (clay) detachment, suspension and re-deposition along the pore wall.[6, 8-10] By adjusting fluid flow or water chemistry, both experimental and analytical research have found fine particle re-arrangement respond to specific factors that govern the particle detachment process.[11-12]

Particle detachment from the rock surface during fluid flow in porous media has been widely studied in the laboratory and also by mathematical modeling.[1, 5, 9, 11, 13-14] The latter being of interest in this study. The analytical model for particle detachment is fundamentally based on a mechanical equilibrium of a particle stationed or attached on a rock surface in a pore channel. Torque balance of all forces acting on a particle is considered, which consists of drag, electrostatic, lifting and gravity forces.[6, 15] Research showed modifying these acting forces could potentially alter particle equilibrium and consequently the amount of retained particles attached on the pore surface.[9, 11, 13, 16-17] Bedrikovetsky *et al*.[6, 8] recently derived a new model to quantify the degree of particle detachment or retention (so-called critical retention concentration) by the torque balance. The authors have matched the model with one-dimensional core-flooding data and found a reasonable agreement under the assumptions of constant filtration coefficient and porosity. Later studies have also demonstrated permeability decline across particle-contained porous cores responding to particle detachment and retention behaviors, which is determined by the critical retention concentration function of the particles.[12, 18-19]

The petroleum reservoir is initially produced by a natural drive, which weakens with increasing production and eventually declines at the end of the primary production. Withdrawal of petroleum fluids and connate formation water leads to an increase in local formation pressure (*i.e.* effective stress) acting upon the reservoir rock underneath (with decreased local pore-pressure to counteract) as generally described in geomechanics.[20-22] Research has shown that increased effective stress can substantially contribute to reservoir deformation and rock matrix micro-cracking.[21, 23-26] Despite a number of experimental and analytical studies on the impact of effective stress, the impact of pore deformation induced by effective stress (*via* reservoir depletion) on particle detachment behavior has yet to be explored. One general assumption of particle detachment models in porous media being a constant porosity has also to consider a stress-dependent effect.[2, 6, 9] As a popular technique attracting industrial and academic interests, low-salinity water injection could also disturb the fines migration phenomenon by altering brine salinity.[12, 14-15, 19, 27-28] Fine (clay) particles hydrated in brine could induce a more hydrated state or "swelling" when brine salinity is decreased.[29-31] With the increase in size brought on by decreasing the brine salinity, both clays attached on the pore surface and suspended in the bulk fluid could potentially change pore configurations and particle detachment/attachment behaviors. For this reason, a holistic study is needed to fully assess the effectiveness of low-salinity EOR.

In the current study, particle detachment is studied analytically based on the critical retention function theory of fine particles. Sensitivity analysis of influencing factors is investigated to evaluate their impacts on fines migration behavior. The analytical model is modified with stress- and salinity-dependent term, aiming to elucidate the relevant consequences for reservoir depletion and low-salinity water injection, respectively. The modified model relaxes the assumption of constant reservoir porosity by incorporating reservoir deformation behavior due to increased effective stress. Reservoir permeability is also determined based on the calculated critical retention concentration of fine particles at various conditions. This demonstrates the influence of particle detachment due to stress- and salinity-dependents on permeability decline.

## 2 Particle Detachment and Critical Retention Concentration Function

In this section, mechanical equilibrium of particle on pore surface is introduced with expressions of all forces acting on the particle. Theory of critical retention concentration function is reviewed and a model estimate is given by assuming a bundle of parallel rectangular capillaries.

### *2.1 Force balance and particle detachment*

According to fines migration theory,[2, 6, 17, 32] each particle (fine) traveling along pore channel is considered with four forces acting on the pore wall (**Figure 1**). The forces consist of (i) drag force ($F_d$), (ii) electrostatic force ($F_e$), (iii) lifting force ($F_l$) and (iv) buoyancy force ($F_g$). The force determinations are summarized below and followed by particle detachment model, which is derived from the force balance. Details of full derivation can be found elsewhere.[6, 33-34]

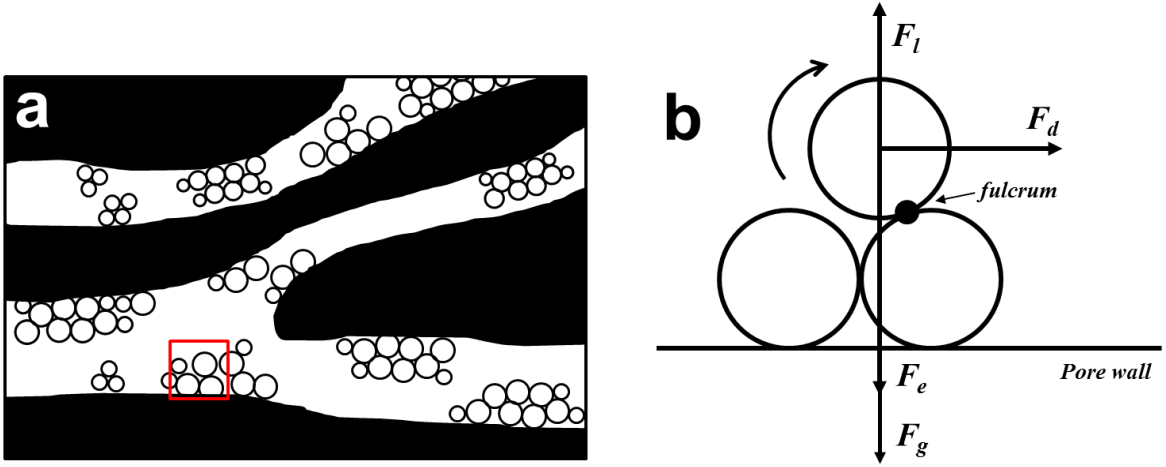

**Figure 1.** Schematics of particle force balance on surface of pore channel within porous media: (a) Particle aggregates as internal cake station on the pore wall surface and (b) Zoom of the red box in (a) shows the forces exerting an individual particle.

**Drag force**: Fluid flow acting on a spherical particle in the Hele-Shaw flow model (*i.e.* flow between two parallel plates) generates the drag force, which is expressed as:[35-36]

$$F_d = \frac{\omega \pi \mu r_s^2 u}{H} \tag{1}$$

where $\omega$ is the drag constant, $\mu$ the continuing fluid (water) viscosity, $r_s$ the equivalent particle diameter, $u$ the average flow velocity and $H$ the pore height.

**Electrostatic force**: Due to the classical DLVO theory, the colloidal force is dominated by an electrostatic force between particle and pore wall. Total colloidal potential ($V$) includes (i) London-van-der-Waals, (ii) double electric layer and (iii) Born potentials:[37-41]

$$F_e = -\frac{\partial V}{\partial h} \tag{2}$$

$$V = V_{LVA} + V_{DLR} + V_{BR} \tag{3}$$

$$V_{LVA} = -\frac{A_{132}}{6}\left[\frac{2(1+Z)}{Z(2+Z)} + ln\left(\frac{Z}{2+Z}\right)\right] \tag{4}$$

$$V_{DLR} = \frac{\varepsilon_0 D_e r_s}{4}\left[2\psi_{01}\psi_{02} ln\left(\frac{1+e^{-\kappa h}}{1-e^{-\kappa h}}\right) - (\psi_{01}^2 + \psi_{02}^2) ln(1-e^{-2\kappa h})\right] \tag{5}$$

$$V_{BR} = \frac{A_{132}}{7560}\left(\frac{\sigma_{LJ}}{r_s}\right)^6\left[\frac{8+Z}{(2+Z)^7} + \frac{6-Z}{Z^7}\right] \quad (6)$$

$$\kappa = \sqrt{\frac{e^2}{\varepsilon_0 D_e k_B T}\gamma} \quad (7)$$

where $Z$ is the ratio of the separation distance between particle and pore wall ($h$) and the particle radius, $A_{132}$ the Hamaker constant, $\psi_{01}$ the particle surface potential, $\psi_{02}$ the collector-particle surface potential, $\varepsilon_0$ the electric constant, $D_e$ the dielectric constant, $\sigma_{LJ}$ the atomic collision diameter, $e$ the elementary charge and $\gamma$ the molar concentration of solution.

**Lifting force**: The force lifting the particle under a constant flow is expressed as[42]

$$F_l = \chi r_s^3 \sqrt{\frac{\rho\mu u^3}{H^3}} \quad (8)$$

where $\chi$ is the lifting constant and $\rho$ the water density.

**Buoyancy force**: The buoyancy force of the particle body is:

$$F_g = \frac{4}{3}\pi r_s^3 \Delta\rho g \quad (9)$$

where $\Delta\rho$ is the density difference between suspended particle and water and $g$ the gravitational constant.

When a particle stations on the pore wall surface, the normal force ($F_n$) reacts upon the particle to balance the force system as:

$$F_n = F_e + F_g - F_l \quad (10)$$

Fines migration occurs in pore space because the resultant force exerts the particle oppositely from the pore wall when a steady-state flow or other dominant factors are changed substantially, *e.g.* increase in flow velocity and decrease in injecting water salinity.[6, 27]

## 2.2 Critical retention concentration function

According to the force balance schematic (**Figure 1**b), the drag and lifting forces detach particles from pore wall surface while the electrostatic and gravity forces attract the particles to the surface. Both drag and lifting forces are function of velocity, while the electrostatic and buoyancy forces are velocity independent. This indicates that the flow velocity is a main factor to release or capture the particles. As a result of particle detachment and its consequent the particle captured or strained into pore space and pore-throat, the pore space and reservoir porosity are reduced accordingly and therefore the interstitial fluid velocity ($U$) increases. The drag force ($F_d$) hence increases with an assumption of constant flow rate.

Considering the last moment when the particle is still stationed or lodged on the rock surface before being released by the drag and lifting forces, at this state the drag force torque is about to exceed the maximum torque of the normal force, which corresponds to electrostatic force maxima. This critical dislodging condition defines a new steady state in pore space after particle detachment and re-deposition. The dimensionless particle dislodging number ($\varepsilon$), which is a ratio between the drag and normal forces, determines the critical or maximum retention concentration ($\sigma$) of the particle as:[6, 33-34]

$$\sigma = \sigma_{cr}(\varepsilon); \; \varepsilon = \frac{\mu r_s^2 |U|}{\sqrt{k_i \phi} F_n} \tag{11}$$

where $\sigma$ is the retention concentration, $\sigma_{cr}$ the critical retention concentration, $\varepsilon$ the particle dislodging number, $U$ the interstitial fluid velocity and $k_i$ the initial permeability.

The critical retention concentration ($\sigma_{cr}$) is a monotonically decreasing function of the particle dislodging number ($\varepsilon$) or mainly the fluid velocity ($U$).[6] The higher the velocity, the higher the drag and the lifting forces resulting in lower critical retention concentration. This implies that an increase in fluid velocity in porous media could substantially induce particle detachment and re-deposition. Low critical retention concentration ($\sigma_{cr}$) refers to high amount of released particles suspending in pore channels and consequently particle re-deposition and pore-plug, which lead to a serious permeability decline.[43]

Bedrikovetsky et al.[6] have estimated the critical retention concentration of particle detachment ($\sigma_{cr}$) as a function of fluid flow velocity ($U$) by assuming a flow in a simple bundle of parallel rectangular capillaries, see **Figure 4**, as.

$$\sigma_{cr}(U) = \left[1 - \left(\frac{\mu r_s^2 U}{\phi H F_e x}\right)^2\right](1 - \phi_c)\phi \tag{12}$$

where the dimensionless parameter $(x)$ is calculated from:

$$1 + \frac{4\pi r_s^3}{3F_e}\Delta\rho g - \frac{\chi\sqrt{\rho F_e}}{\mu} x^{3/2} = \sqrt{3}\omega\pi x \tag{13}$$

The dimensionless parameter $(x)$ is derived from the force balance of Eq. (10) with Eqs. (1), (2), (8) and (9) in order to simplify Eq. (12). The full details and related discussions can be found in Bedrikovetsky *et al*.[6] It is noted that a flow in a bundle of capillaries presumes no tortuosity ($\tau = 1$) in order to simplify the model formulation as widely assumed in the established publications,[6, 44-46] even though there is always tortuosity at some degree in porous media ($\tau < 1$). It is also well known that flow paths with high tortuosity are considered to be severe plugging pathways with serious throat-blocking results,[47] while the current study diminishes such variation and rather considers the effects of other parameters as will be discussed in Section 3.

## 3 Particle Detachment under an Effective Stress Influence

### *3.1 Deformation of reservoir configurations due to an increasing effective stress*

After oil has been produced from natural drive (*i.e.* potential energy from subsurface) in primary recovery, reservoir fluid (pore) pressure gradually decreases due to oil depletion. Decrease in pore pressure promotes an increase in effective stress over reservoir formation since the total stress (bulk pressure) is constant. This is described by Terzaghi's principle:[20]

$$P' = P_t - P_p \tag{14}$$

where $P'$ is the effective stress, $P_t$ the total stress (due to bulk formation) and $P_p$ the pore pressure.

Increase in an effective stress results in compaction on reservoir formation and closure of micro-channels which consequently change in reservoir configurations, thus flow path tortuosity is modified while reservoir permeability and porosity altered.[21, 23, 48-50] Ma *et al*.[23] have studied change in porosity of core sample as a result of an increase in effective stress using core-flooding technique. Confining pressure was applied on core samples as total stress at the same overburden pressure in the field that core samples were extracted. As shown in **Figure 2**, core sample porosity decreased with increasing effective stress. As stress is applied (< 35 MPa), porosity changes significantly as it responds to the new conditions. With increasing stress values (> 35 MPa), the stress impact on porosity gradually declines. This is because the compressed micro-cracks and open throats do not crack or are not compressed further once a certain degree of stress burden was reached.[21, 25-26, 51-52] The relationship between core porosity and effective stress can be logarithmically correlated as:

$$\frac{\phi}{\phi_i} = \varphi - a\ln(P') \tag{15}$$

where $\phi$ is the formation porosity and $\phi_i$ the initial formation porosity.

The regression parameter ($a$) is a fitting parameter from the experimental data while the stress-formation constant ($\varphi$) is defined as a unique parameter for a specific formation-stress behavior. Based on Eq. (15), the constant $\varphi$ can be determined from the initial condition at $\left(P', \frac{\phi}{\phi_i}\right) = (P'_i, 1)$:

$$1 = \varphi - a\ln(P'_i) \tag{16}$$

where $P'_i$ is the initial effective stress.

It is noted that the initial effective stress cannot be zero ($P'_i \neq 0$) in an actual reservoir since the petroleum formation stations underground with certain and substantial overburden from ground surface. It is noted that the stress-formation constants ($\varphi$) in both cases in **Figure 2** are similar because they were determined from the same-source core samples and applied the same stress set in the experiment.

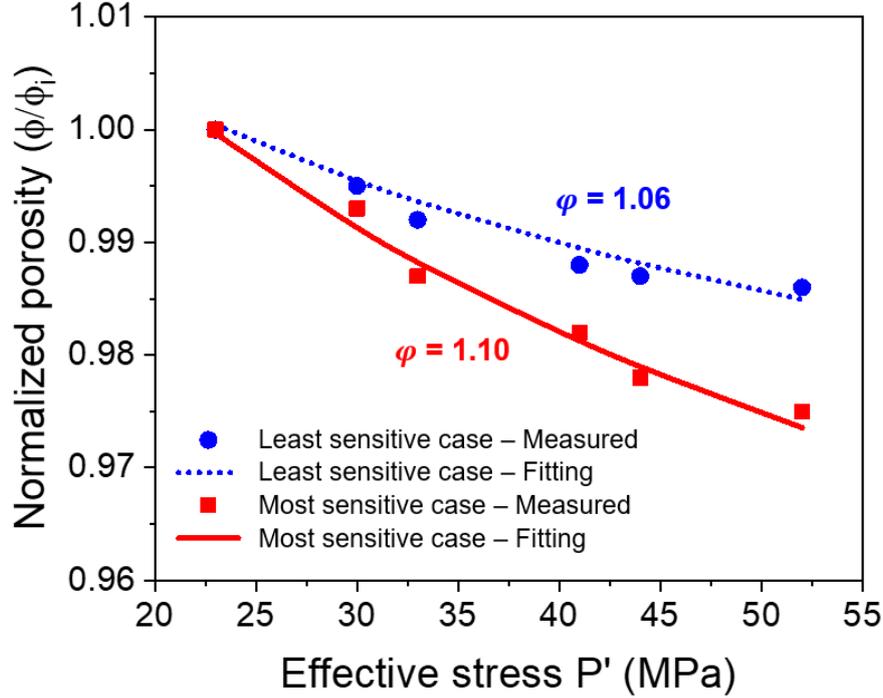

**Figure 2.** Decrease in reservoir porosity with increasing effective stress. Experimental results by Ma *et al.*[23] (symbols) can be correlated by logarithmic function (lines) with a stress-formation constant ($\varphi$). With the influence of increased effective stress, the most sensitive case is the case that porosity was decreased relatively at the highest degree while the least sensitive case is the case that porosity was decreased relatively at the lowest degree.

Likewise, an internal cake formed by residual and/or suspended particles depositing on pore wall inside permeable rock are also compressed and deformed due to an impact of effective stress under the same stress environment to the hosting rock (**Figure 3**). The compression of the internal cake thus behaves similarly to the reservoir porosity, see Eq. (15).[53-54] The cake porosity is expressed as a function of an effective pressure as:

$$\frac{\phi_c}{\phi_{ci}} = \varphi_c - a_c ln(P') \qquad (17)$$

where $\phi_c$ is the cake porosity, $\phi_{ci}$ the initial cake porosity and $a_c$ the cake regression parameter.

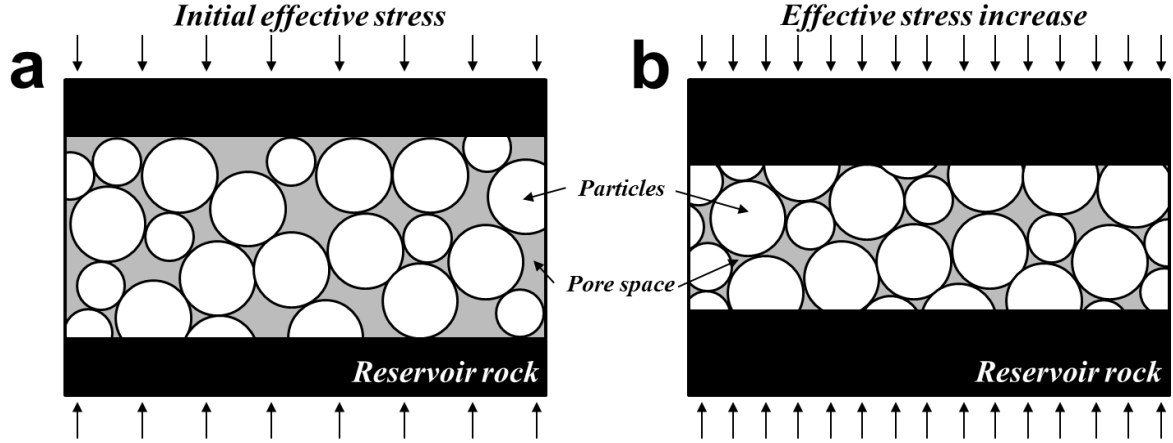

**Figure 3.** Schematics of cake and reservoir configuration change due to increased effective stress. Similar to reservoir porosity, initial cake porosity at initial effective stress (a) decreases with increasing effective stress (b).

The cake stress-formation constant ($\varphi_c$) is a stress-formation parameter for a specific internal cake which theoretically differ from the reservoir stress-formation constant ($\varphi$). However, since both reservoir porosity and cake porosity are compressed under the same stress environment, their deformations could be assumed to behave similarly and have to have the same stress-formation constant ($\varphi \approx \varphi_c$).

Furthermore, a construct of effective stress also inevitably impacts on pore sectional areas of permeable flow paths in porous rock, which could then disturb a flow behavior/regime substantially (*e.g.* increase fluid flow velocity) and need to be considered.

To simplify a solution, pore sectional area or pore space of porous media is considered as a bunch of parallel rectangular tubes (*i.e.* the Hele-Shaw flow[55], see **Figure 4**).[6, 56-57] With pore concentration ($n$) and pore opening size ($H$), the porosity of porous rock can then be determined:[58-59]

$$\phi = nH^2 \tag{18}$$

Assuming pore concentration ($n$) is not changed after porous rock has been compressed due to increasing stress, the pore opening size deformation can be defined *via* Eq. (15) as:

$$\frac{H}{H_i} = \sqrt{\varphi - a\ln(P')} \qquad (19)$$

where $H_i$ is the initial pore opening size.

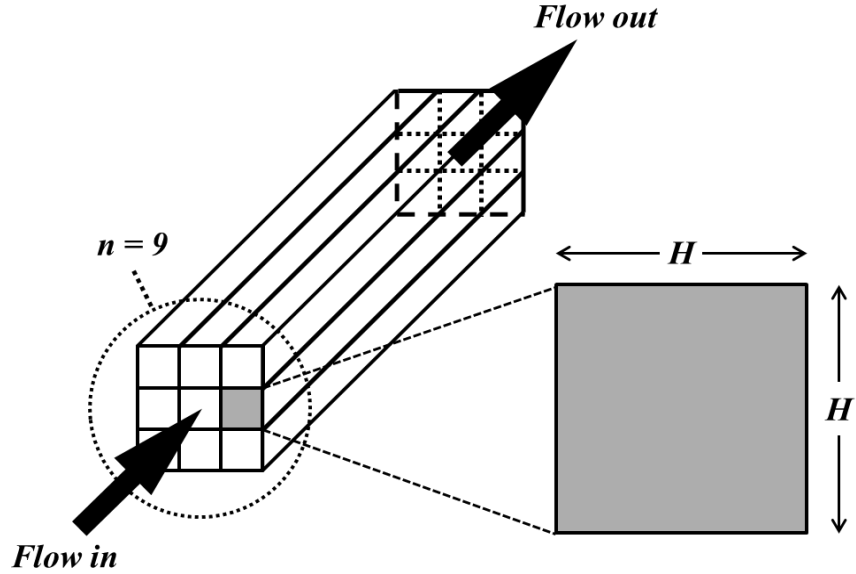

**Figure 4.** Schematic of the Hele-Shaw flow with a bunch of parallel rectangular pores showing the pore opening size ($H$) and pore concentration (*i.e.* $n = 9$).

*3.2 Critical retention concentration as a stress-dependent function*

According to the previous model,[6] the flow rate ($q$) in porous reservoir is assumed to be constant and can be determined by cross-sectional area ($A$) of a rectangular capillary:

$$q = AU = H^2 U \qquad (20)$$

Substituting the normalized fluid velocity ($U/U_i$) in Eq. (19), the interstitial fluid velocity ($U_i$) is then accordingly changed due to increased effective stress with the stress parameters as:

$$\frac{U}{U_i} = \frac{1}{\varphi - a\ln(P')} \qquad (21)$$

Considering the critical retention concentration function, reservoir configuration parameters ($\phi, \phi_c$ and $H$) and fluid flow velocity ($U$) in Eq. (12) are then replaced by the parameters that are impacted by effective stress. Reservoir porosity ($\phi$) is a stress-dependent parameter relaxing a typical assumption of constant porosity.[2, 6, 9] The particle detachment behavior is therefore expressed as a new stress-dependent function of critical retention concentration ($\sigma_{cr}(P')$):

$$\sigma_{cr}(P') = \left[1 - \frac{1}{M^5}\left(\frac{\mu r_s^2 U_i}{\phi_i H_i F_e x}\right)^2\right](1 - \phi_c M_c)\phi M_c \qquad (22)$$

where $M$ is $\varphi - a\ln(P')$ and $M_c$ is $\varphi_c - a_c \ln(P')$.

Fluid flow velocity ($U$) in the new particle detachment model is also a function of an effective stress (Eq. (21)), therefore this new model directly expresses the critical retention concentration as a function of an effective stress ($P'$). The other reservoir parameters are also considered in the model, which includes reservoir porosity ($\phi$) affected by an effective stress. It is noted that such change in reservoir porosity relaxes an assumption of the conventional model.[6, 11, 60] The new model was validated with published experimental data showed in **Supplementary Material**, where a good agreement is found.

*3.3 Sensitivity analysis and discussion*

To elucidate an influence of effective stress on particle detachment, the modified critical retention concentration function is examined through sensitivity analysis on significant parameters including the initial fluid velocity ($U_i$), particle size ($r_s$), lifting constant ($\chi$), and drag constant ($\omega$).[6] The stress-formation parameters (**Table 1**) define reservoir characterizations under the effective stress using Eqs. (15), (17), (19) and (21). In this study, the cake stress-formation constant is assumed to be 98% of the reservoirs stress-formation constant. The electrostatic force is calculated from factors reported in **Table 2**, following Bedrikovetsky *et al*.[6] The dimensionless parameter $x$ is then determined from Eq. (13) by reservoir properties and initial reservoir configurations shown in **Table 3**, including the parameters of the base case in sensitivity analysis. The base-case parameters are estimated from reported literature to best describe typical formation damage behavior, especially the cake porosity.[3, 6, 32] Varying the selected sensitivity parameters ($U_i, r_s, \chi$ and $\omega$) ranging from 50% to 150% of the base-case value, the dimensionless parameter $x$ is changed accordingly and the critical retention concentration ($\sigma_{cr}$) is then calculated.

**Table 1.** Stress-formation parameters for sensitivity analysis.[2, 17]

| Parameter | Value |
|---|---|
| Stress-formation constant ($\varphi$) | 1.080 |
| Cake stress-formation constant ($\varphi_c$) | 1.058 ($\approx 0.98\varphi$) |
| Regression parameter ($a$) | 0.027 |
| Cake regression parameter ($a_c$) | 0.019 |

**Table 2.** Parameters for electrostatic force ($F_e$) calculation.

| Parameter | Value | Ref. |
|---|---|---|
| Hamaker constant ($A_{132}$) | $2 \times 10^{-21}$ J | 41 |
| Surface potential of particle ($\psi_{01}$) | $-30$ mV | 40 |
| Surface potential of collectors-grains ($\psi_{02}$) | $-50$ mV | 40 |
| Electric constant ($\varepsilon_0$) | $8.854 \times 10^{-12}$ C$^{-2}$J$^{-1}$m$^{-1}$ | 40 |
| Dielectric constant for water ($D_e$) | 78.0 | 40 |
| Atomic collision diameter in Lennard-Jones potential ($\sigma_{LJ}$) | 0.5 nm | 40 |
| Solution salinity ($\gamma$) | 0.513 mM | 6 |
| Electrolyte valence for sodium chloride ($z$) | 1 | 6 |

**Table 3.** Reservoir and model parameters (base case) for sensitivity analysis.

| Parameter | Value |
|---|---|
| Initial porosity ($\phi_i$) | 0.186 |
| Initial cake porosity ($\phi_{ci}$) | 0.100 |
| Initial permeability ($k_i$) | 118 mD |
| Initial fluid velocity ($U_i$) | 0.0001 m/s |
| Particle size ($r_s$) | 0.442 μm |
| Lifting constant ($\chi$) | 1,190 |
| Drag constant ($\omega$) | 60 |
| Pore opening size ($H$) | 3.99 μm |
| Water viscosity ($\mu$) | 1 cp |
| Water density ($\rho$) | 1,000 kg/m$^3$ |

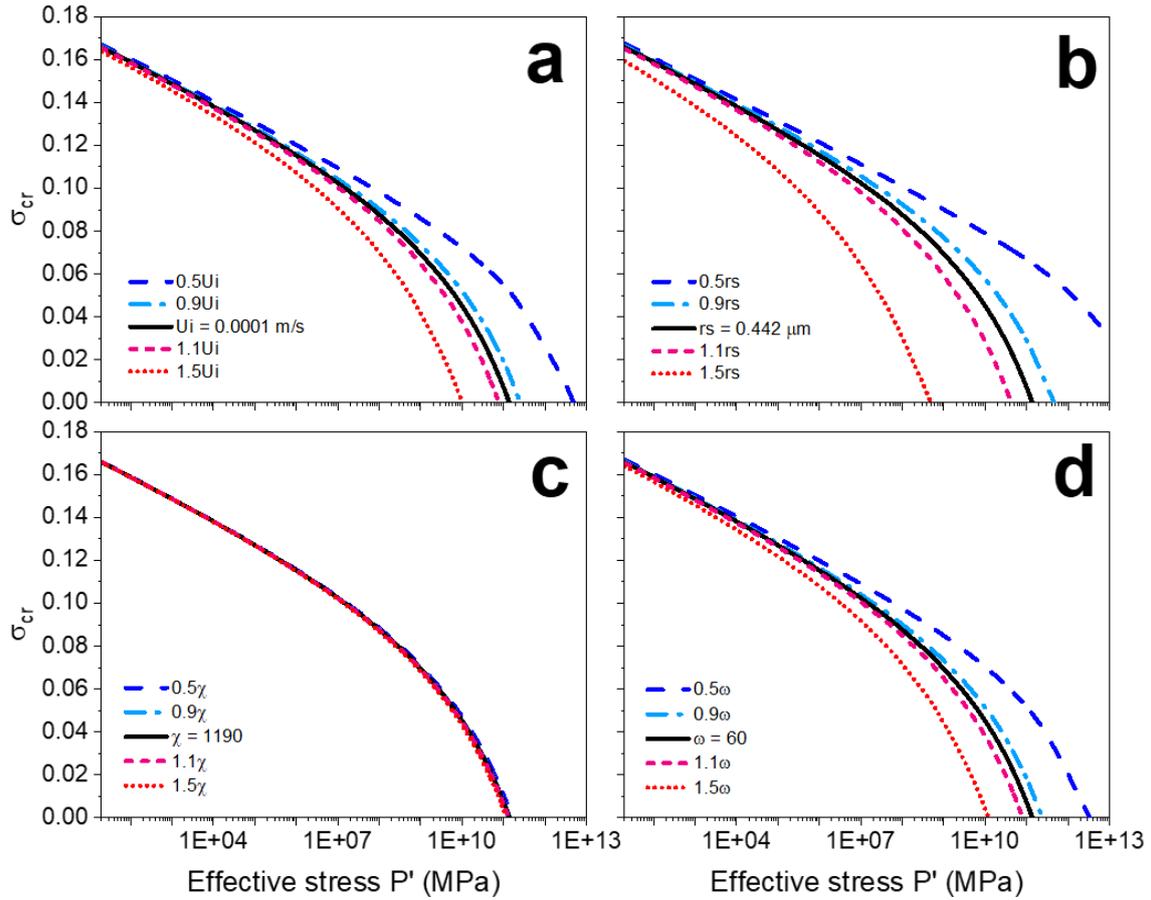

**Figure 5.** Sensitivity analysis on critical retention concentration as a function of effective stress: with varying (a) initial fluid velocity ($U_i$); (b) particle size ($r_s$); (c) lifting constant ($\chi$); and (d) drag constant ($\omega$). Base cases are shown in solid lines while the cases with varying sensitivity are shown in dash lines as indicated.

**Figure 5** shows analytical results of the critical retention concentration with the four varying parameters. All studied cases result in the same trend, the critical retention concentration is inversely proportional to an effective stress. The results imply that the particles are increasingly suspended and re-deposit further with increasing effective stress (*i.e.* increasing reservoir depletion). Decrease in $\sigma_{cr}$ could also be attributed to the effect of reservoir compaction and hence reduced porethroats due to an increasing effective stress as studied by Haghi *et al.*[21] Reduced critical retention concentration at high effective stress could potentially lead to serious permeability decline within the porous media and thus low injectivity of displacing fluids.[11, 18]

As the original model suggests,[6] the initial fluid velocity ($U_i$) has a remarkable influence on particle detachment. Significant differences in the initial fluid velocity, as shown in **Figure 5**a, emphasizes that the critical retention concentration decreases with increasing fluid velocity. High fluid velocity promotes drag and lifting forces acting on particles which results in high particle mobilization, hence less particles are retained or captured. The critical retention concentration also decreases with increasing particle size (**Figure 5**b), as larger fines particles are subjected to stronger drag force and increased repulsive electrostatic force. Increase in drag force is due to increased particle surface area. Also, with larger particles, electrostatic interaction between particle and wall surface is enhanced as expressed by the Derjaguin approximation.[61] It should be also noted that the effect on the electrostatic force is slightly higher than that of the drag force but the two are in the same order of magnitude, suggesting a slight higher influence of the electrostatic force on the particle detachment than the drag force. Interestingly, the particle size ($r_s$) has a greater effect on its migration than the fluid velocity, while the critical retention concentration is much more sensitive especially where $P' > 10^9$ MPa. For the largest particle size ($r_s = 0.663$ μm) at $P' > 10^9$ MPa, the analysis shows $\sigma_{cr} = 0$. This is evidence that all particles become detached and suspended in pore channel, which is large enough ($H$ decreased from 3.99 to 2.88 μm at $P' = 10^9$ MPa) to accommodate the particles ($H > 2r_s$). This also elucidates a significance of fine particle (clay) uniqueness over the flow mechanics of fluid. It is important to consider the effecting factors that influence the fine particle size (*e.g.* water chemistry or salinity) in the particle detachment model to better understand the fines migration behavior and to achieve less formation damage.

On the contrary, there is no substantial effect of the lifting constant ($\chi$) on disturbing particle detachment and thus the critical retention concentration (**Figure 5**c). All sensitivity analysis for the lifting constant results in similar $\sigma_{cr}$ function over the studied range of effective stress. This could be due to the lifting force formula (Eq. (8)), which derives from the lifting constant to the 1$^{st}$ ($\chi$) while the particle size to the 3$^{rd}$ ($r_s^3$). Thus, the particle size serves to nullify the lifting constant impact. Sensitivity of the drag constant ($\omega$) directly corresponds to the drag force formula (Eq. (1)), which promotes particle detachment and thus reduced critical retention concentration via increased drag force. Analysis results show the drag constant is as sensitive as fluid velocity ($U$) but less sensitive than the particle size ($r_s$), see **Figure 5**d in relation to the impact on critical retention concentration.

As the most sensitive and significant parameter regarding fines migration, the fine particle (clay) size is of interest in the next section where it can be altered by change in water salinity due to swelling/de-swelling properties. Changed or modified injecting water salinity is also relevant to low-salinity water injection, which is widely reported as one of EOR technique.

# 4 Particle Detachment due to Clay Swelling during Low-Salinity Water Injection

*4.1 Clay particle swelling due to low-salinity water injection*

Decrease in salinity of continuous water phase causes dispersed-clay swelling due to osmotic pressure and cation exchanges depending on clay minerals and mineral crystalline structure. When brine contacts clay minerals and cation surface charge of clay interlayer is greater than salt ions, water will migrate into the space between clay layers and hence clay interlayer swells.[29] Montmorillonite clay has a higher ability to swell due to its very high cation exchange capacity compared to kaolinite.[30] Zhuang *et al*.[31] have reported that montmorillonite clay swelled from ~3.9 to 5.1 µm diameter (~30%) after brine salinity (NaCl) decreased from 5,000 mg/L (0.09 mM) to pure water. Fang *et al*.[62] have observed the clay swelling impact on permeability decline with ~38% decreasing across clay-containing core sample after changing injection fluid from formation water (9,662.9 mg/L) to pure water. Clay swelling could cause a serious formation damage since detached clays are swelled and suspended to re-deposit into pore spaces. This leads to reduced pore area or blocked fluid flow paths, thus resulting in permeability decline. The swelling or expansion of clay is defined by a swelling ratio ($\zeta$), which is always higher than 1.

$$\zeta = \frac{V(\gamma)}{V} = \left(\frac{r_s(\gamma)}{r_s}\right)^2 \tag{23}$$

where $V(\gamma)$ is the clay volume that swells due to salinity decrease to $\gamma$ salinity and $V$ the initial clay volume. By assuming clay as spherical particle, the clay particle radius ($r_s$) will expand to be $r_s(\gamma)$ at $\gamma$ salinity.

## 4.2 Critical retention concentration with clay swelling-dependent function

Low-salinity water injection would potentially induce a swelling of attached clay on pore surface, which consequently influences flow channel dimensions and pore configurations. Reservoir parameters are changed by considering a swelling ratio. Thus, the stress-dependent critical retention concentration of the remaining captured particles can be determined with the new parameters. It is noted that only physical effect of clay swelling (due to change in brine salinity) is examined in this section in order to elucidate an impact of clay swelling *per se* on fines migration behavior, not including a concurrent effect of change in electrostatic force as also a result of change in salinity.

The initial pore opening area ($H_i^2$) is reduced by subtracting excess swelling area of clay particle ($r_s^2(\gamma) - r_s^2$) due to decrease in water salinity. This assumes one layer of clay particles. The initial pore opening size with clay swelling considered ($H_i(\gamma)$) is thus:

$$H_i^2(\gamma) = H_i^2 - [r_s^2(\gamma) - r_s^2] = H_i^2 - r_s^2(\zeta - 1)$$

$$H_i(\gamma) = \sqrt{H_i^2 - r_s^2(\zeta - 1)} \tag{24}$$

Assuming the pore concentration ($n$) is constant after swelling and the reservoir and cake porosities are affected by change in salinity in a similar manner, the reservoir porosity and the cake porosity are then reduced as:

$$n = \frac{\phi_i}{H_i^2} = \frac{\phi_i(\gamma)}{H_i(\gamma)^2} \tag{25}$$

$$\phi_i(\gamma) = \phi_i \left(\frac{H_i(\gamma)}{H_i}\right)^2 \tag{26}$$

$$\phi_{ci}(\gamma) = \phi_{ci} \left(\frac{H_i(\gamma)}{H_i}\right)^2 \tag{27}$$

The interstitial fluid velocity ($U_i$) from Eq. (20) is also changed to be $U_i(\gamma)$ with a constant flow rate:

$$U_i(\gamma) = U_i \left(\frac{H_i}{H_i(\gamma)}\right)^2 \tag{28}$$

In summary, the parameters that are changed according to salinity decrease consist of the clay particle size ($r_s$), initial pore opening size ($H_i$), initial reservoir porosity ($\phi_i$), initial cake porosity ($\phi_{ci}$) and initial fluid velocity ($U_i$). Thus, the stress-dependent particle detachment model (Eq. (22)) can be integrated with low-salinity effect as follows:

$$\sigma_{cr}(P',\gamma) = \left[1 - \frac{1}{M^5}\left(\frac{\mu r_s(\gamma)^2 U_i(\gamma)}{\phi_i(\gamma)H_i(\gamma)F_e x(\gamma)}\right)^2\right](1-\phi_c(\gamma)M_c)\phi(\gamma)M_c \qquad (29)$$

Here, the parameter $x(\gamma)$ is calculated from Eq. (13) with parameters that consider an effect of particle swelling.

*4.3 Sensitivity analysis and discussion*

Combined influence of effective stress and injecting brine salinity was examined to study the particle detachment behavior using the modified model of Eq. (29). In the analysis, degree of clay swelling as a function of brine salinity was extracted from the study by Zhuang *et al.*[31] The brine salinity decreases from 5,000 to 0 mg/L (0.09 to 0 mM NaCl) and the associated swelling ratios are shown in **Table 4**. By using the same stress-formation parameters (**Table 1**) and initial reservoir properties (**Table 3**) in the previous section, the reservoir parameters altered due to clay swelling are calculated and then used to determine the critical retention concentration ($\sigma_{cr}(P',\gamma)$). It is noted that changes in electrostatic force due to salinity in this analysis are negligible, since the calculated parameter $x$ has no significant difference.[63] Sensitivity analysis of the particle (clay) initial size ($r_s$) in brine salinity ranging from 5,000 to 0 mg/L is shown in **Figure 6**.

Table 4. Clay swelling ratio due to brine salinity decrease (5,000 to 0 mg/L).

| Brine salinity (mg/L) | Swelling ratio |
|---|---|
| 5,000 | - |
| 2,500 | 1.33 |
| 1,000 | 1.19 |
| 500 | 1.04 |
| 0 | 1.04 |

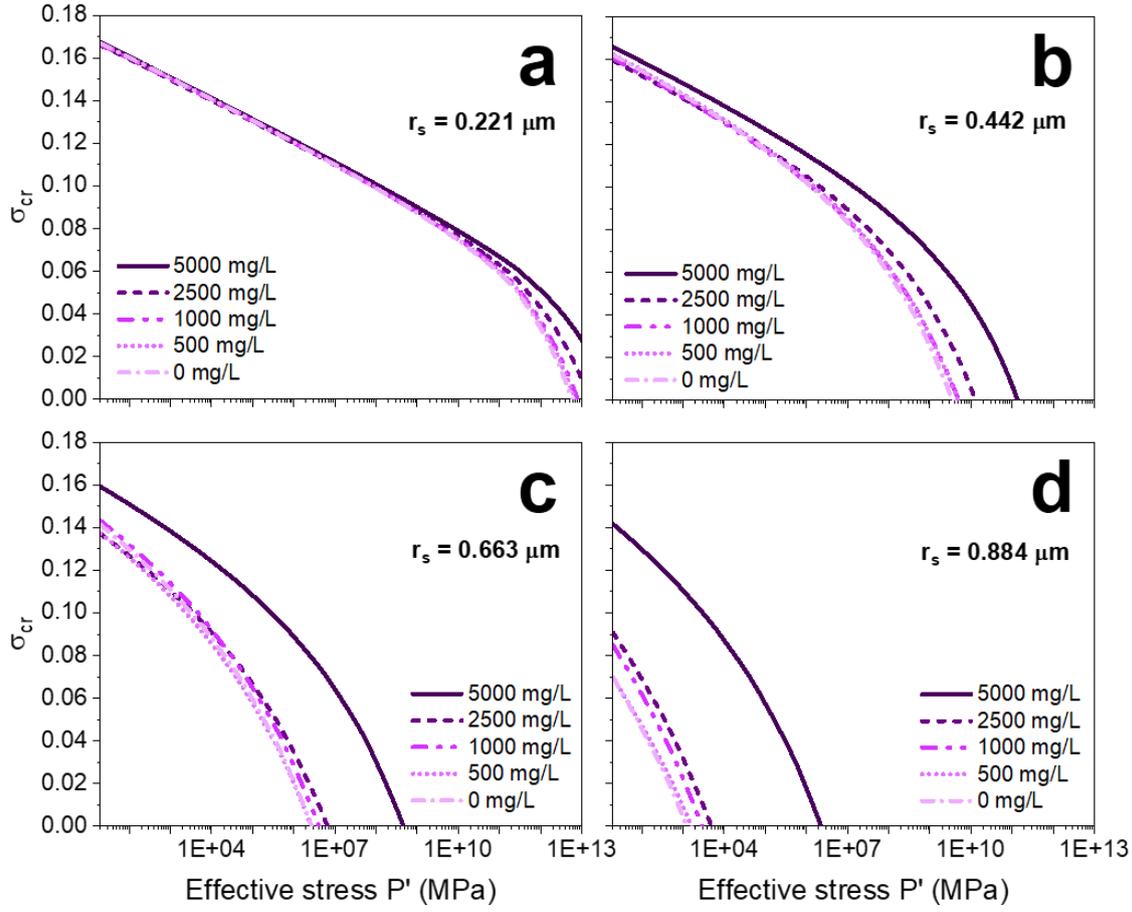

**Figure 6.** Sensitivity analysis on the critical retention concentration as a function of an effective stress and brine salinity. Fine particle initial radii are (a) 0.221, (b) 0.442, (c) 0.663 and (d) 0.884 µm.

As shown in **Figure 6**, the critical retention concentration as a function of salinity still develops the same trend as in the previous section, where no salinity-dependence is considered. This trend sees the $\sigma_{cr}(P',\gamma)$ decrease with increasing effective stress. Change in salinity is seen to have an effect on the $\sigma_{cr}(P',\gamma)$ corresponding to the clay particle size ($r_s$). The small clay particle size (**Figure 6**a) results in minimal change to the $\sigma_{cr}(P',\gamma)$, whereas the larger clay particles (*e.g.* **Figure 6**d) found a significant change with the $\sigma_{cr}(P',\gamma)$ significantly decreasing with decreasing salinity. According to the force balance that governs the particle detachment, difference in the $\sigma_{cr}(P',\gamma)$ is solely due to the change in particle size ($r_s$) and pore height ($H$) rather than enhancing or reducing detaching forces (*i.e.* drag, lifting and buoyancy). Only these two variations in Eqs. (1), (8) and (9) are directly affected by the particle swelling and consequently promote the particle detachment by increasing those forces. The plots in **Figure 7** show a significant increase in constituent terms (*i.e.* the product of $r_s$ and $H$) of the drag, lifting and buoyancy forces for large particle sizes (*e.g.* $r_s$ = 0.884 µm) while a smaller particle ($r_s$ = 0.221 µm) is found to contribute negligibly. This confirms the dominant role particle size plays in directly influencing component forces and subsequently particle detachment and retention. In addition, with very small particle sizes (**Figure 6**a), micro and sub-micro pore spaces are not fully plugged by clay particles leading to marginal detaching forces, this suggests a weak particle suspension with possibly no re-arrangement. The particles are likely retained, resulting in only a slight change in the $\sigma_{cr}(P',\gamma)$ and possibly negligible permeability decline.[62, 64-65]

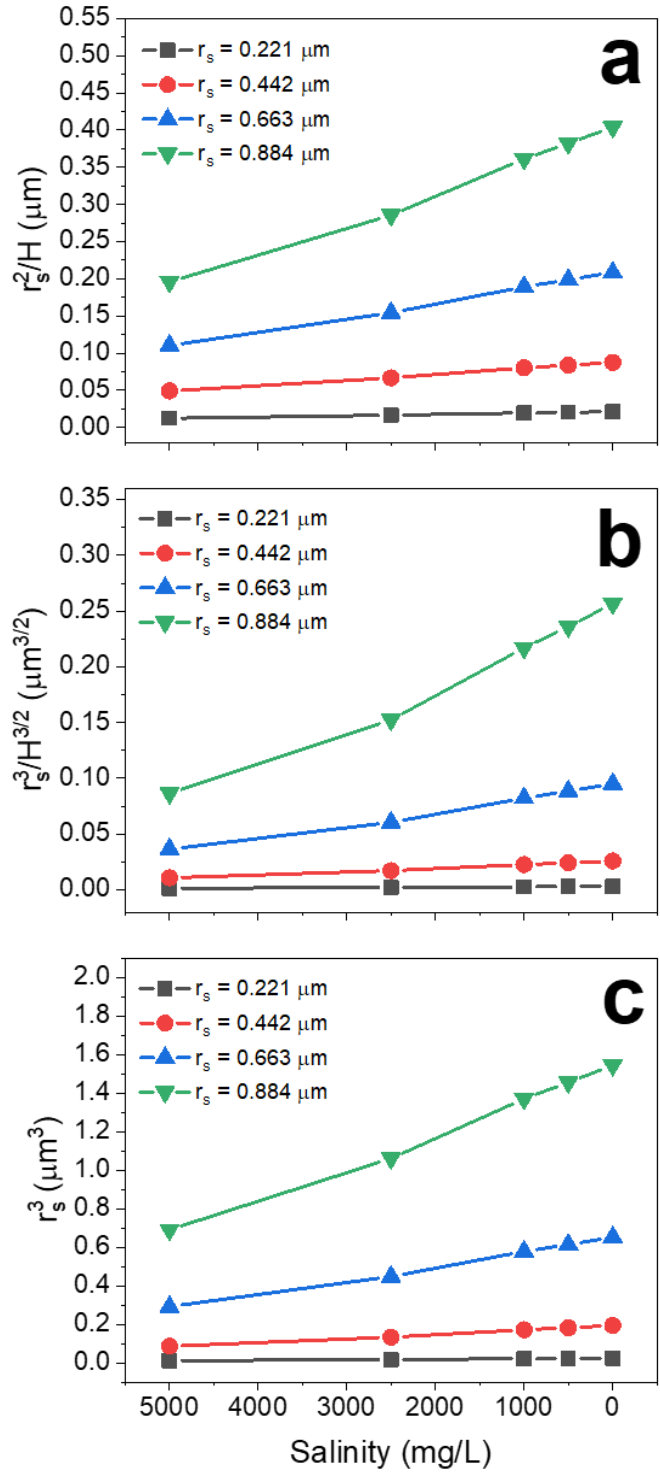

**Figure 7.** Constituent terms (product of $r_s$ and $H$) of the drag (a), lifting (b) and buoyancy (c) forces as a function of brine salinity and initial clay particle size. Lines to guide the eye.

As clay swells greater at more diluted brine concentrations this implies the promotion of particle detachment, with decreases in brine salinity resulting in lower critical retention concentration. It is noted that the lower critical retention concentration in this section was only as a result of physical clay swelling, since the electrostatic force would have changed negligibly within the studied range of salinity (5,000 to 0 mg/L). It is generally considered in literature that diluting brines detach the attached clay particles due to weaker electrostatic attractive forces,[16, 18, 63, 66] However such phenomenon can be negligible and has not been considered in the present section due to very low salinity (< 1 mM NaCl).

Interestingly, dilution in brine salinity has a distinct influence on clay particle detachment up to a certain concentration (*i.e*. 2,500 mg/L, see **Figure 6**) where further dilutions result in approximately the same particle retention concentration even though the detaching forces tend to increase (**Figure 7**). This occurs as the swelling property of the clay becomes relatively minimal at $\gamma \leq 1{,}000$ mg/L (**Table 4**) and the pore space has already decreased to such an extent that it obstructs particle travelling through, thus particle mobilization is reduced. Additionally, further dilution in brine salinity (2,500 to 0 mg/L) has no effect on particle detachment (*i.e*. implying no significant permeability decline). This could help determine the optimum salinity range that is suitable for typical low-salinity injection techniques in order to anticipate other low-salinity EOR mechanisms, *e.g.* wettability alteration and the oil-water interfacial tension reduction.[67-72]

Furthermore, potential impacts of crude oil polarity in regard to fines migration should be addressed, although the present work solely examines an effect of single-phase flow. Since it has been shown that low-salinity technique could influence in both wettability alteration (*via* weakening crude oil-rock electrostatic force and thus reducing attractive disjoining pressure between oil-water and rock-water surfaces[73-75]) and fines migration,[4] recent studies have deliberately excluded the wettability effect from the system in order to elucidate a sole effect of fines migration by using non-polar oil that has no electrostatic impact.[76-77] The studies yet observed considerable additional oil recovery when oil-saturated porous media was flooded with low-salinity brines. As expected, additional oil was produced with large volume of fines, confirming attribution of fines mobilization to incremental oil recovery. In regard to polar oil, its contribution on fines migration is also crucial. The polar components in crude oil favorably interact or 'age' with fine particles, thus the particles are shielded by oil while the particle-brine interaction is hindered. This could result in a reduction in particle detachment or fines migration.

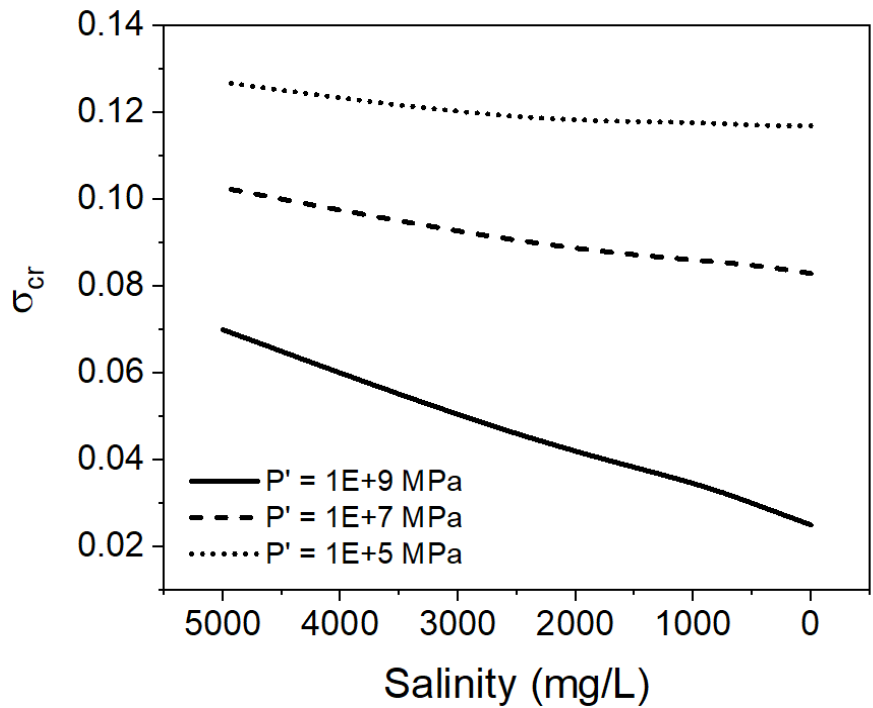

**Figure 8.** Decrease in critical retention concentration with decreasing brine salinity at various effective pressures ($r_s = 0.442$ μm).

The influence of effective stress on critical retention concentration was re-plotted as a function of salinity (**Figure 8**) for 0.442 μm particle sizes. Results shows that the effective stress influence on the $\sigma_{cr}(P',\gamma)$ screens the brine salinity effect, especially at low effective stress. Change in $P'$ from $1 \times 10^5$ to $1 \times 10^9$ MPa resulted in ~80% decrease in the $\sigma_{cr}(P',\gamma)$ while only ~7% decrease was found after brine dilution from 5000 to 0 mg/L at $P' = 1 \times 10^5$ MPa. However, at high effective stress (*e.g.* $1 \times 10^9$ MPa) decreasing salinity increasingly influences the $\sigma_{cr}(P',\gamma)$ since particle detachment is greater due to coupled contributions of the two factors. This implies that petroleum reservoirs subject to highly applied effective stresses (*e.g.* deep reservoir) would have more severe formation damage due to particle detachment once the low-salinity water injection is introduced.[78] In this case, a holistic study needs to assess whether low-salinity EOR technique will enhance oil production or induce formation damage within the reservoir.

## 5 Discussion on Permeability Decline

After some of particles detached, the initial particle concentration in pore space could simply divided into two groups: (i) the retained particles defined by the critical retention concentration function ($\sigma_{cr}$) and (ii) the rest of particles flowing along pore channel which is later strained ($\sigma_s$) or plugged in porous media and causing permeability decline.[27] It is assumed that all particles detached will be later strained in porous media, although in fact detached particles might be not all strained but some suspended or mobilized further. This can form a simple relation as:

$$1 = \sigma_{cr} + \sigma_s \tag{30}$$

Reservoir permeability depends on the strained concentration as described by:[79-80]

$$\frac{k(\sigma)}{k_0} = \frac{1}{1+\beta\sigma_s} = \frac{1}{1+\beta(1-\sigma_{cr})} \tag{31}$$

where $k_0$ is the initial reservoir permeability, $k(\sigma)$ permeability due to particle retention ($\sigma$) after formation damage and $\beta$ formation damage coefficient.

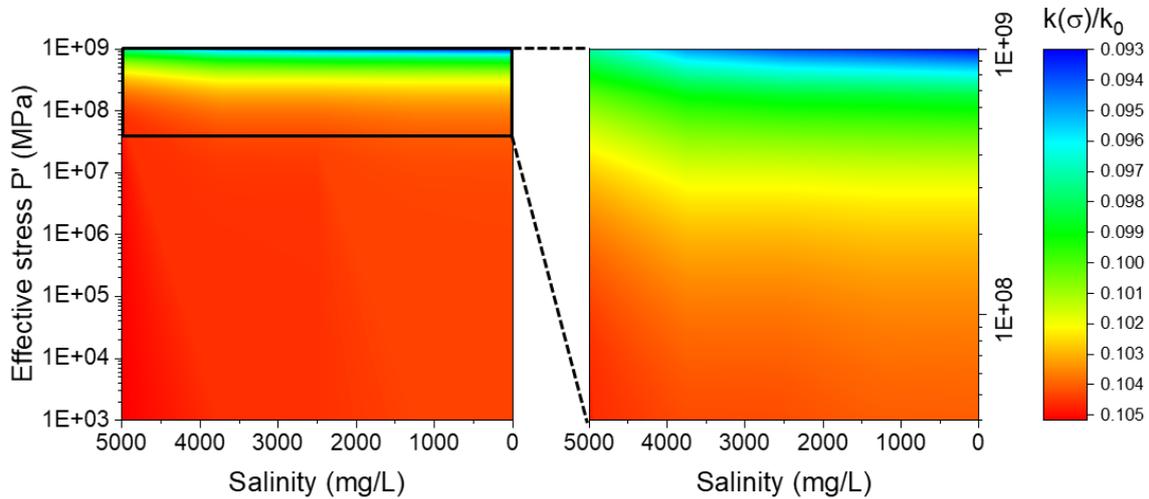

**Figure 9.** Permeability decline ($k(\sigma)/k_0$) as a function of effective stress and brine salinity shown in contour map. Right map is a zoom of effective stress ranging $4 \times 10^8$ - $1 \times 10^9$ MPa.

Permeability decline defined by normalized permeability with an initial permeability ($k(\sigma)/k_0$) was calculated in the ranges of effective stress and brine salinity of the previous section with $r_s = 0.884$ µm and $\beta = 10$.[14] Contour map shown in **Figure 9** indicates higher sensitivity of brine salinity found at higher effective stress ($> 1 \times 10^8$ MPa), while at effective stress $< 1 \times 10^8$ MPa the analytical results obtains likely the same ($k(\sigma)/k_0$ ~0.1). Effective stress is more dominant than brine salinity on permeability decline according to particle detachment and deposition phenomena. The higher the effective stress, the higher the permeability decline. At the effective stress $\geq 1 \times 10^8$ MPa, decrease in brine salinity increasingly contributes to reduced permeability and implies the formation damage. Significant permeability decline at high effective stress and low brine salinity is a consequence of particle detachment in the porous media as discussed by sensitivity analysis in the previous section. The results from this analysis have a good agreement with reported experimental observations,[12, 27, 81] where particle detachment could cause serious consequence the pressure drop between injection and production wells and hence the injectivity decline.

With increasing depth in permeable reservoir, effective stress continually increases and hence induces particle detachment and retention. This generates different degree of fines migration which directly results in effective fluid flow gradient and changed flow tortuosity. The effective permeability therefore decreases from near ground surface towards deeper level into the reservoir (*i.e.* fines-induced formation damage).[19, 82] More severe permeability decline would occur at the deepest layer of reservoir while the shallower layer would have less fines migration, thus reservoir fluid prefers to flow effectively in the shallower layer. This phenomenon could then construct a kind of "fines-assist slug" driving or controlling a fluid flow convection, which is undesirable because the average reservoir permeability is decreased. Such problem caused by distinctive effective pressure gradient *per se* is likely inevitable for the massive and deep reservoirs. However, alteration in brine salinity could potentially weaken or reduce the "undesired flow convection" by deploying the "fines-assisted low-salinity waterflood".[19, 83] As a result of an analytical study, the present work (**Figure 9**) shows a great sensitive of brine salinity on permeability and yet further optimization is needed.

# 6 Conclusions

Critical retention concentration function describing fine particle detachment in porous media has been modified based on the model by Bedrikovetsky *et al.*[6] with effective stress and brine salinity dependents considered. As reservoir pore-pressure continually depletes, the effective stress increases and compresses reservoir pore configurations, which induces more particle detachment. Brine salinity controls particle size *via* clay swelling property, which also promotes particle block at fluid flow paths with decreasing salinity. Based on analytical study and sensitivity analysis, the present work can be concluded as follows:

1. Under influence of effective stress, particle size and fluid velocity are found to be remarked dominant factors controlling the fines migration. Increase in particle size or fluid velocity directly promotes the detaching forces that exert the attached particles to detach and suspend.

2. With decreasing brine salinity, clay particles that attached on pore surface could increasingly swell and directly reduce effective pore area and flow channel dimensions. Such phenomenon potentially promotes particle blocking the fluid flow paths and is proven to cause serious permeability decline.

3. Initial particle size of clay is main factor governing particle detachment behavior. Small clays are likely not detached from pore surface due to weak detaching forces. This suggests unchanged critical retention concentration and negligible permeability decline.

4. Integrated analysis of both stress- and salinity-dependent on fines migration finds that salinity-dependent effect is screened by stress-dependent effect, especially at low effective stress. However, the salinity-dependent effect increases to dominate at higher effective stress.

**Competing interests**

The authors declare no competing interests.

**Acknowledgement**

The authors would like to thank Mr. Ngurah Beni Setiawan (Imperial College London, UK) for valuable discussion and express their gratitude to the reviewers who help to improve the quality of the work.

**Nomenclature**

| | |
|---|---|
| $a$ | regression parameter |
| $a_c$ | cake regression parameter |
| $A$ | cross-sectional area, μm$^2$ |
| $A_{132}$ | Hamaker constant, J |
| $D_e$ | dielectric constant |
| $e$ | elementary charge, C |
| $F_d$ | drag force, N |
| $F_e$ | electrostatic force, N |
| $F_g$ | buoyancy force, N |
| $F_l$ | lifting force, N |
| $F_n$ | normal force, N |
| $g$ | gravitational constant, m/s$^2$ |
| $H$ | pore opening size, μm |
| $H_i$ | initial pore opening size (height of the channel), μm |
| $H_i(\gamma)$ | initial pore opening size with clay swelling considered, μm |
| $k_0$ | initial reservoir permeability, mD |
| $k_i$ | initial permeability, mD |
| $k(\sigma)$ | permeability due to the particle retention, mD |
| $M$ | stress-formation grouping |
| $M_c$ | cake stress-formation grouping parameters |
| $n$ | pore concentration |
| $P'$ | effective stress, MPa |
| $P'_i$ | initial effective stress, MPa |
| $P_p$ | pore pressure, MPa |
| $P_t$ | total stress due to bulk formation stress from surface to reservoir, MPa |
| $q$ | flow rate, m³/s |
| $r_s$ | particle size (equivalent diameter), μm |
| $r_s(\gamma)$ | particle size at $\gamma$ salinity, μm |
| $U$ | interstitial fluid velocity, m/s |
| $U_i$ | initial interstitial fluid velocity, m/s |
| $U_i(\gamma)$ | initial at $\gamma$ salinity, m/s |
| $u$ | average flow velocity at a given cross-section area, m/s |
| $V$ | total potential energy, J |
| $V_{BR}$ | Born potential energy, J |
| $V_{DLR}$ | double electric layer potential energy, J |
| $V_{LVA}$ | London-van-der-Waals potential energy, J |
| $x$ | dimensionless parameter |
| $x(\gamma)$ | dimensionless parameter at $\gamma$ salinity |
| $Z$ | ratio of the separation distance between particle and pore wall |
| $V$ | initial clay volume, μm³ |
| $V(\gamma)$ | clay volume that swells due to salinity decrease to $\gamma$ salinity, μm³ |

*Greek Letters*

| | |
|---|---|
| $\beta$ | formation damage coefficient |
| $\gamma$ | molar concentration of solution, mM |
| $\varepsilon_0$ | electric constant, C$^{-2}$J$^{-1}$m$^{-1}$ |
| $\varepsilon$ | particle dislodging number |
| $\zeta$ | swelling ratio |
| $\kappa$ | Debye length, m$^{-1}$ |
| $\mu$ | water viscosity, cp |
| $\rho$ | density of water, kg/m$^3$ |
| $\Delta\rho$ | density difference between the suspended particles and water, kg/m$^3$ |
| $\sigma$ | particle retention concentration |
| $\sigma_{cr}$ | particle critical retention concentration |
| $\sigma_{LJ}$ | atomic collision diameter, nm |
| $\sigma_s$ | particle strained concentration |
| $\tau$ | tortuosity |
| $\phi$ | formation porosity |
| $\phi_c$ | cake porosity |
| $\phi_c(\gamma)$ | cake porosity at $\gamma$ salinity |
| $\phi_{ci}$ | initial cake porosity |
| $\phi_i$ | initial formation porosity |
| $\phi_i(\gamma)$ | initial formation porosity at $\gamma$ salinity |
| $\varphi$ | stress-formation factor |
| $\varphi_c$ | cake stress-formation factor |
| $\chi$ | lifting coefficient |
| $\psi_{01}$ | particle surface potential, mV |
| $\psi_{02}$ | collector-particle surface potential, mV |
| $\omega$ | drag constant |